\documentclass[a4paper]{jpconf}
\usepackage{graphicx}
\usepackage{floatflt}
\begin{document}

\newcommand{\snn}{\sqrt{s_{NN}}}
\newcommand{\seff}{\s_{\rm eff}}
\newcommand{\s}{\sqrt{s}}
\newcommand{\AAA}{A+A}
\newcommand{\pp}{p+p}
\newcommand{\pbar}{\overline{p}}
\newcommand{\pbarp}{\overline{p}+p}
\newcommand{\qbarq}{\overline{q}+q}
\newcommand{\epem}{e^+e^-}

\newcommand{\nhit}{N_{hit}}
\newcommand{\npp}{n_{pp}}
\newcommand{\nch}{N_{ch}}
\newcommand{\np}{N_{part}}
\newcommand{\ntot}{\langle\nch\rangle}
\newcommand{\avenp}{\langle\np\rangle}
\newcommand{\npB}{N_{part}^B}
\newcommand{\nc}{N_{coll}}
\newcommand{\avenc}{\langle\nc\rangle}
\newcommand{\half}{\frac{1}{2}}
\newcommand{\halfnp}{\langle\np/2\rangle}
\newcommand{\etap}{\eta^{\prime}}
\newcommand{\as}{\alpha_{s}(s)}
\newcommand{\etazero}{\eta = 0}
\newcommand{\etaone}{|\eta| < 1}
\newcommand{\dndeta}{d\nch/d\eta}
\newcommand{\dndetap}{d\nch/d\etap}
\newcommand{\dndetazero}{\dndeta|_{\etazero}}
\newcommand{\dndetaone}{\dndeta|_{\etaone}}
\newcommand{\dndetanp}{\dndeta / \halfnp}
\newcommand{\dndetapnp}{\dndetap / \halfnp}
\newcommand{\dndetaonp}{\dndeta / \np}
\newcommand{\dndetazeronp}{\dndetazero / \halfnp}
\newcommand{\dndetaonenp}{\dndetaone / \halfnp}
\newcommand{\ratio}{\ntot/\halfnp}
\newcommand{\nee}{N_{ee}}
\newcommand{\nhh}{N_{hh}}
\newcommand{\nubar}{\overline{\nu}}
\newcommand{\yb}{y_{\rm beam}}
\newcommand{\mpt}{\langle p_T \rangle}

% add words to TeX's hyphenation exception list
\hyphenation{author another created financial paper re-commend-ed Post-Script thermo-dyna-mic-ally}

\title{Hotter, Denser, Faster, Smaller...and Nearly-Perfect: What's the matter at RHIC?}

\author{Peter Steinberg}

\address{Brookhaven National Laboratory, Upton, NY 11973}

\ead{peter.steinberg@bnl.gov}

\begin{abstract}
The experimental and theoretical status of the ``near perfect fluid'' at 
RHIC is discussed.  While the hydrodynamic paradigm for understanding 
collisions at RHIC is well-established, there remain many important open
questions to address in order to understand its relevance and scope.
It is also a crucial issue to understand how the early equilibration is 
achieved, requiring insight into the active degrees of freedom at
early times.

\end{abstract}

\section{Introduction}
In April 2005, the four experiments at RHIC made a joint announcment
of the discovery of a ``perfect liquid'' in high energy 
Au+Au 
collisions~\cite{RHICrelease,Arsene:2004fa,Adcox:2004mh,Back:2004je,Adams:2005dq}.
This discovery has made quite an impression
in the popular press, resulting in articles in Scientific
American (April 2006)
and Discover magazine (February 2007).  It even was called the 
most important scientific story of 2005 by the American
Institute of Physics~\cite{AIP}.  While the press release was somewhat
short on details, it offered a tantalizing glimpse of the
connections with subjects as far afield as black holes
and string theory, suggesting that they offered possible
insight into the strongly-coupled system formed at RHIC.

The foundation of these discoveries was the success of ideal
hydrodynamics in modeling the experimental data in a robust
way.  In other words, the system formed at RHIC turns out to
be best seen as a single system that evolves collectively,
rather than an ensemble of individual nucleon-nucleon collisions.
The goal of this review is to outline the techniques and
successes to date of hydrodynamics at RHIC, discuss the possible
experimental and theoretical limits of this interpretation,
and show how the current data points to new horizons in both
regimes.

\section{The ``Perfect Fluid'' at RHIC}

\begin{floatingfigure}[r]{80mm}
\begin{center}
\includegraphics[width=65mm]{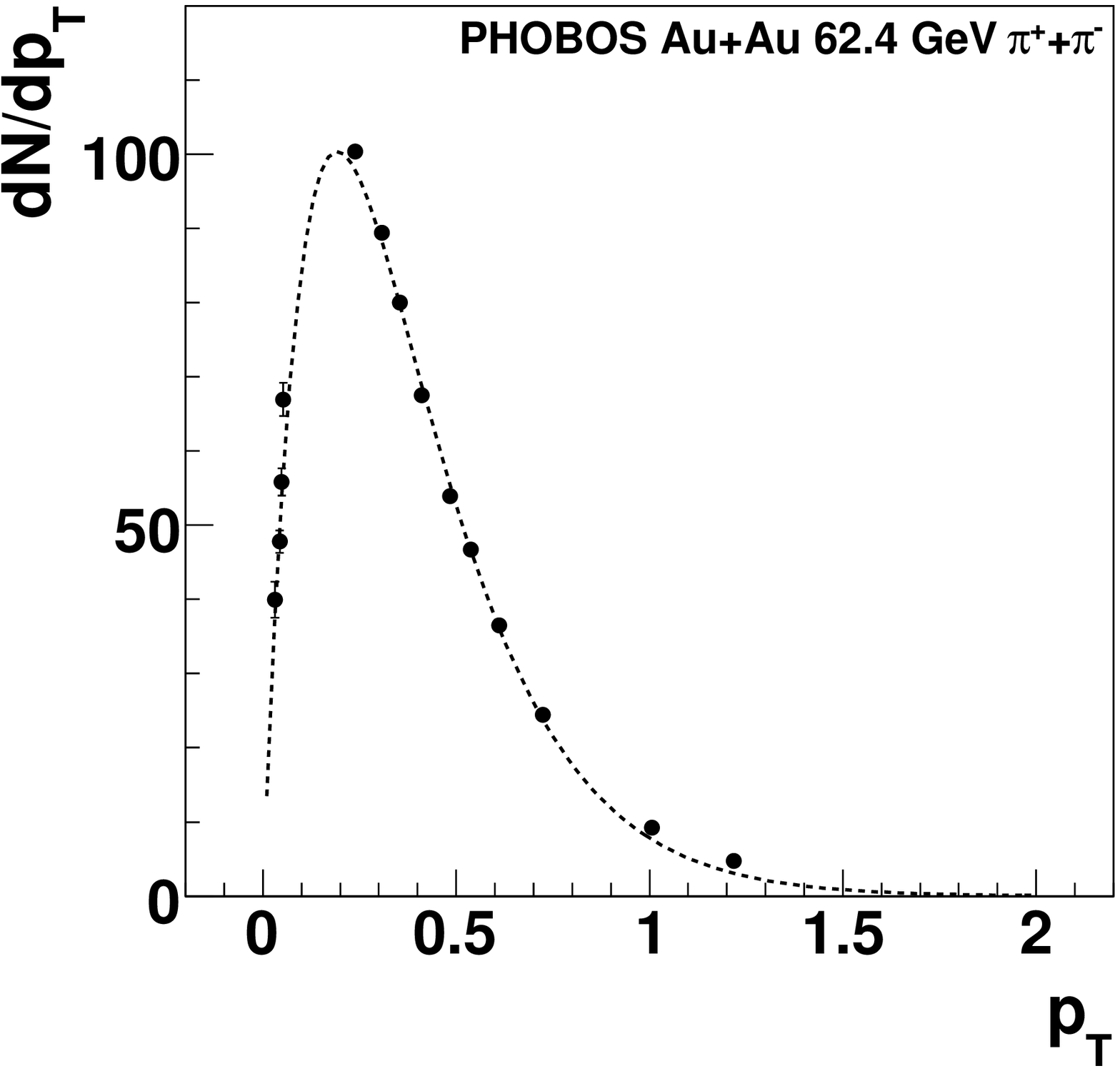}
\caption{
PHOBOS $dN/dp_T$ for charged pions at  
$\snn=62.4$ GeV.  A Boltzmann fit is shown.
\label{fig:blackbody}}
\end{center}
\end{floatingfigure}

In hydrodynamical calculations for RHIC physics,
there are typically three basic 
stages~\cite{Landau:gs,Kolb:2003dz}: initial conditions,
hydrodynamic evolution, and freeze-out.  The initial conditions
consist of defining the energy density in position space, with
each fluid cell given an initial velocity (note that these are
purely classical calculations!).  The aspects most interesting
for RHIC collisions are when the spatial distribution of the
initial energy density is highly anisotropic.  This is because
the second stage of the collision takes place while the system
remains in local thermal equilibrium and evolves according to
the evolution equations of hydrodynamics:
\begin{equation}
\partial_\mu T^{\mu \nu} = 0
\end{equation}
which are closed by the equation of state
%\begin{equation}
$p = f(\epsilon)$
%\end{equation}
e.g. $p=\epsilon/3$ for a gas of photons or QCD at high
temperatures~\cite{Boyd:1996bx}.  During this period,
gradients in pressure and energy density induce
an expansion and associated cooling of the system.
As the local energy density decreases, so does the temperature
($T \propto \epsilon^{1/4}$).  When a fluid element cools to the scale of
the Compton wavelength of the pion $T_{ch} \sim m_\pi$, the
system is considered to be sufficiently dilute that it decouples
or ``freezes-out'' of the evolution and is transformed into hadrons.
This occurs as an isotropic decay of the available energy density
into known hadrons with thermal abundances in the fluid rest 
frame~\cite{Belenkij:cd,Cooper:1974ak}.

While this physical scenario was invented in the early 1950's and
studied through the 1970's, it was eventually discarded as a good
model for strong interactions.  Many rejected the possibility
that small systems had sufficient time (or, strong enough coupling
of the different fluid elements) to achieve any sort of local
thermal or chemical equilibrium, much less a kinetic equilibirium
of the final state particles, as suggested by Hagedorn in
the 1960's~\cite{Hagedorn:1965st} based on the exponential rise
of the hadron mass spectrum.%~\cite{Yao:2006px}.
Heavy ion collisions are larger and generate higher multiplicities,
leading to a general expectation that equilibrium is likely.

In RHIC collisions, circumstantial hints of early-time thermalization
shows up in the relative population of various hadron states, both
as a function of particle mass and transverse momentum.
The transverse momentum dependence, shown in Fig.~\ref{fig:blackbody} 
from Ref.~\cite{Back:2006tt} fit by a Boltzmann distribution, 
has the familiar
form of the blackbody spectrum:
$f(p,m)=1/(\exp[(\sqrt{p^2+m^2}-\mu_B)/T]\pm 1)$
although it is obviously a ``strong blackbody''
with hadrons in the place of photons.  

\begin{floatingfigure}{70mm}
\begin{center}
\includegraphics[width=65mm]{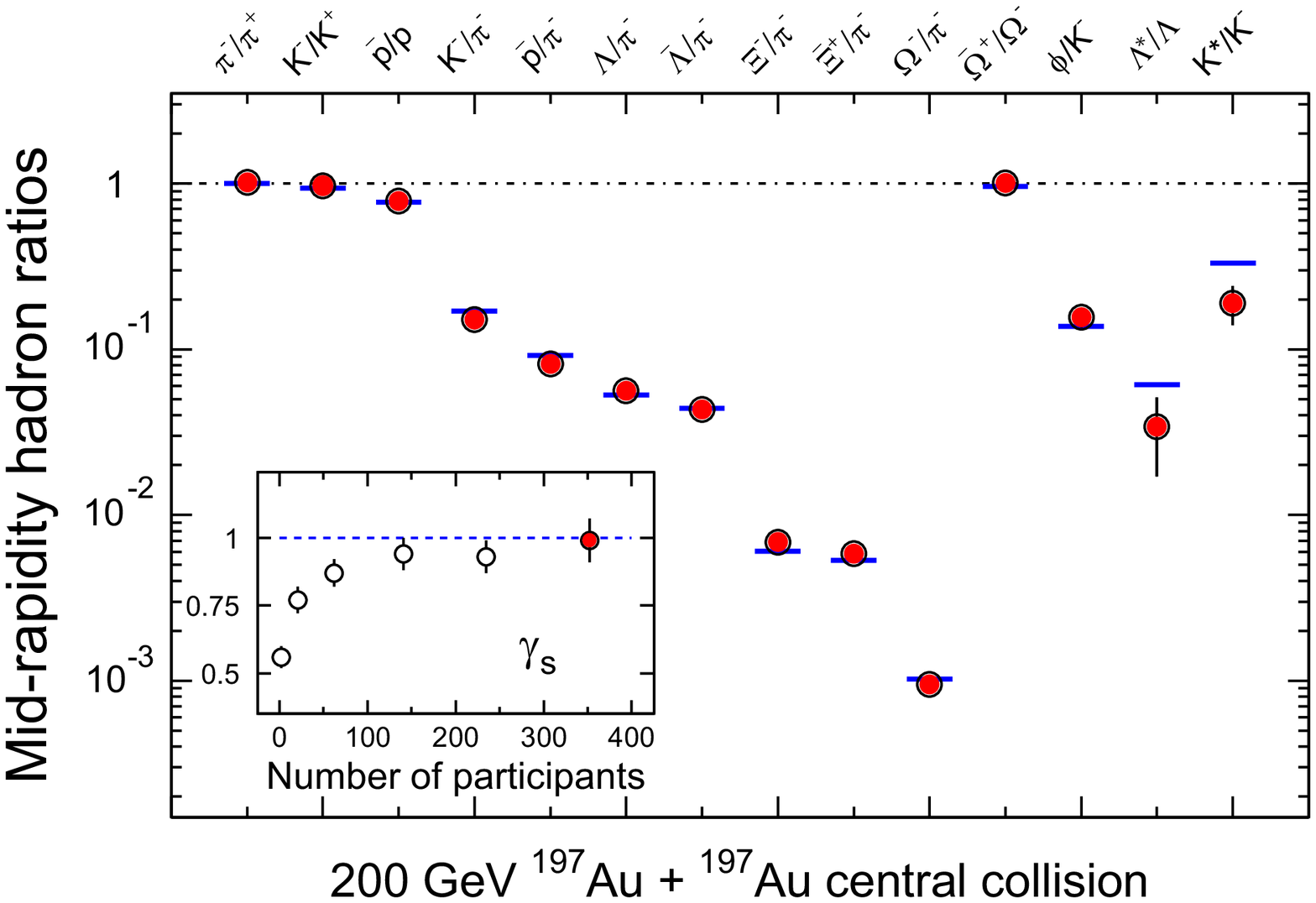}
\caption{
(left) STAR data on hadron ratios fit to a thermal model, from
Ref.~\cite{Adams:2005dq}.
\label{fig:thermalmodel}}
\end{center}
\end{floatingfigure}

When the spectrum of each particle species is integrated
$N(m)\propto V \int d^3 p f(p,m)$, it is conventional to take
ratios of particle yields to cancel out the common volume factor.
This simple ``thermal model''~\cite{Braun-Munzinger:2003zd,Cleymans:1998fq}
depends primarily on two parameters,
the freeze-out $T_{ch}$ and the baryon chemical potential 
$\mu_B$ (defined such that $\overline{p}/p \sim exp(-2\mu_B/T_{ch})$).  
The experimental data on hadron yields is fit to a 
model incorporating these paramters (and all the known strong
decays of high mass resonances to the observed particles) to
estimate $T_{ch}$ and $\mu_B$.  Typical values from recent
fits are $T_{ch} = 177$ MeV~\cite{Adams:2005dq}, corresponding to 
$2 \times 10^{12} K$.  According to the hydrodynamical scenario
outlined above, this suggests that all possible hadron states
are available to the system as it freezes out, so it may well make
sense to assume that the relevant degrees of freedom active 
{\it before} the freezeout were also in thermal and chemical
equilibrium.  This itself implies that the system was substantially
hotter than $T_{ch}$ for most of the evolution, especially at
the very beginning.  Thus, $10^{12}$ degrees is the {\it coolest} the
system can be!  It should be noted, however, that the relevance
of a single temperature does not imply that the entire system freezes
out in some sort of ``global'' equilibrium.  Rather, this temperature
seems to reflect
the local properties of the system induced by the properties of
the known hadron spectrum.%~\cite{Yao:2006px}.

\begin{floatingfigure}[r]{80mm}
\begin{center}
\includegraphics[width=50mm]{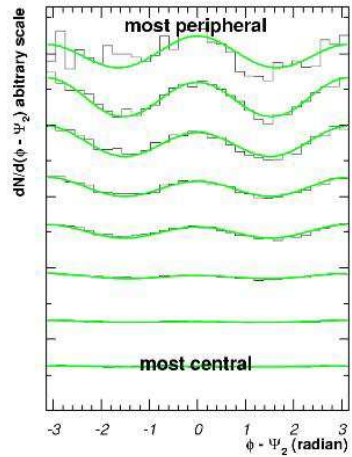}
\caption{
Charged $dN/d\phi$ relative to the event plane for
for different centralities in
$\snn = 200$ GeV Au+Au collisions from PHOBOS.
\label{fig:phobos-flow}}
\end{center}
\end{floatingfigure}

If the system is in local equilibrium just before freezeout,
then it is certainly plausible that it could show hydrodynamic
behavior.   Hydrodynamics implies the relevance of
thermodynamics, but of course not the converse does not necessarily
hold true~\cite{Belenkij:cd}.
At RHIC (and other machines colliding nuclei) it has been noticed
that the event-by-event angular distribution is not isotropic
in azimuthal angle.  An ``event plane'' can be estimated from the 
produced particles, defined as angle $\Psi_R$ of 
the long principal axis of the particle angles.
The azimuthal distribution relative to this event plane is found
to show a strong $\cos(2[\phi-\Psi_R])$ dependence, shown 
in Fig.~\ref{fig:phobos-flow}, from Ref.~\cite{Back:2002gz}.  This is
especially pronounced in more peripheral (i.e.\ lower multiplicity)
collisions, where the overlap of the nuclei is shaped like an
almond, relative to central (i.e.\ higher multiplicity) collisions
where the overlap is essentially isotropic.  This leads to
a characterization of the event-by-event angular distributions in
terms of its Fourier coefficients~\cite{Voloshin:1994mz}:
\begin{equation}
\frac{dN}{d\phi} = 1+2v_1 \cos(\phi-\Psi_R) + 2v_2 \cos(2[\phi-\Psi_R]) + \cdots
\end{equation}

\begin{floatingfigure}[l]{80mm}
\begin{center}
\includegraphics[width=75mm]{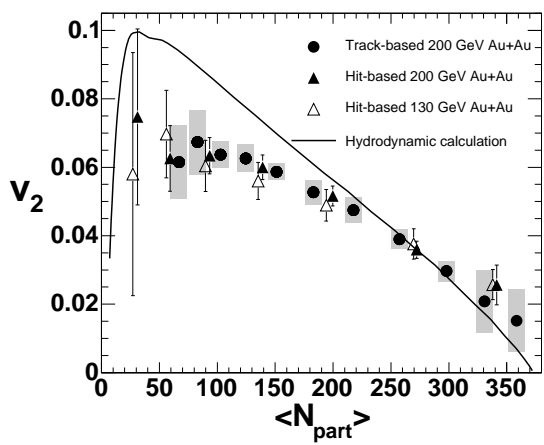}
\caption{
The elliptic flow parameter $v_2$ vs. $\np$ for $\snn = 200$ GeV
from PHOBOS compared with a hydrodynamical model.
\label{fig:WP6}}
\end{center}
\end{floatingfigure}

One of the early and striking results from RHIC was the measurement
of $v_2$ -- conventionally called ``elliptic flow'' --
as a function of centrality (e.g. the number of participating
nucleons $\np$), an example of which is shown in Fig.~\ref{fig:WP6}
from Ref.~\cite{Back:2004mh}.  
The data show the qualitative features described
above, that peripheral collisions have a large value of $v_2$, reflecting
the large asymmetry in the initial distribution of nucleons, while
the central collisions have a much smaller value.  Hydrodynamic calculations
use estimates of various properties (e.g. energy or entropy density) of 
the initial conditions by extrapolating from lower-energy data, and 
turn out to compare well to the data for 
$N_{part}>200$~\cite{Kolb:2000fh}.  At lower
energies, similar calculations often overpredicted $v_2$ by 
factors of two, so RHIC data is considered striking validation of the
hydrodynamic approach.

Comparing the relevant energy and space-time scales implied by the
success of the hydrodynamical models, the matter at RHIC is formed
under quite extreme conditions.  The estimates of the
formation time relevant for the hydrodynamic calculations
were predicted to be in the vicinity of $\tau_0 = 0.6$ fm/c, or 
approximately 
2 yoctoseconds ($10^{-24}$)\cite{Kolb:2000fh}.  
This number is far smaller than the
time taken a massless particle to traverse the radius of a hadron
($\tau \sim 1$ fm/c)~\cite{Bjorken:1982qr}.
At this time, the energy density needed to match the data is around
$\epsilon = 30$ GeV/fm$^3$.  This should be compared with the energy
density of a nucleon in its rest frame, $\epsilon_N \sim 500$ MeV/fm$^3$,
which it exceeds by a factor of 60.  And in the same sense as
$T_{ch}$ is a lower limit of the system temperature at early times,
these estimates do not preclude even higher energy densities at
even earlier times.  All of this depends on exactly when the system
can be said to be in local thermal equilibrium, i.e.\ on the
precise value of $\tau_0$.

Thus, to summarize, the success of the hydrodynamic models at RHIC
suggest that collisions there make something that 
is hotter, denser, smaller, and faster (in 
the sense of thermalization time) than other known liquids.  
Furthermore, ideal hydrodynamics is inviscid by construction, i.e.\
there are no non-equilibrium processes encoded by
the equations, at least until freezeout.
In this light, the appelation of ``perfect fluid'' seems quite reasonable.

\section{The Edge of Liquidity: The ``Near Perfect Fluid''}

\begin{floatingfigure}{70mm}
\begin{center}
\includegraphics[width=65mm]{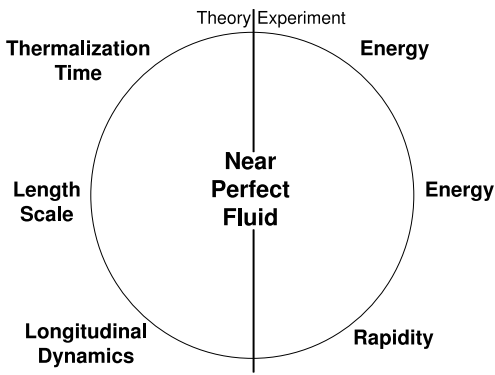}
\caption{
\label{fig:frontiers}
Frontiers of the hydrodynamic paradigm
}
\end{center}
\end{floatingfigure}

However, it does not take much serious inquiry to raise a few
doubts on this conclusion.  No viscosity was needed to 
reproduce the data within experimental and theoretical uncertainties.
And yet, it is not obvious that incorporating viscosity to some level would
make the models completely disagree with the experimental data.
Thus, it may be a ``nearly'' perfect liquid instead of a merely
perfect liquid.
But what exactly characterizes a ``liquid'', as opposed to a strongly
coupled gas?  Both are fluids, while liquids are distinguished by
attractive interactions between the constituents that generate a
surface tension.  Of course some sort of attractive interactions generate the
bound-state hadrons in the final state, but this says nothing
about earlier times.

Thus, until the precise form of the interaction can be
shown to be uniformly attractive, and until it can be demonstrated that
the system evinces no viscosity, the appellation ``perfect liquid'' should
probably be replaced by the ``near perfect fluid''.  
This does not reduce the importance of the RHIC discoveries.  Rather 
it points to future investigations in both experiment and theory
to elucidate the detailed microscopic properties of this system.
%It is interesting to note that both experiment and theory can continue
%to make strides independent of one another.  Eventually it will
%be the responsibility of phenomenologically-minded theorists and
%experimentalists to make the connection between the two groups
%on the basis of experimental data and models based on theoretical
%principles.
In the meantime, we will assume the near-perfect fluid interpretation
as a paradigm for looking at the RHIC data.  The idea will be to
vary certain parameters in experiment and theory, to see if any
major changes take place that are qualitatively different from
what is observed in the most-central collisions.

In order to probe the limits of this near-perfect fluid paradigm
one can push to the ``edge of liquidity'' (shown schematically
in Fig.\ref{fig:frontiers} experimentally
(in energy, geometry, or rapidity) or theoretically 
(in thermalization time, length scale, or longitudinal dynamics)
to see if the qualitative behavior changes.
One way to examine the roles energy and geometry simultaneously,
we can consider the systematics of elliptic flow simultaneously 
as a function of energy, centrality and system size.  
In the past 15 years, a large
data set on $v_2$ has been measured near $\eta=0$ in the center-of-mass
frame as a function of energy and 
centrality~\cite{Adcox:2004mh,Back:2004je,Adams:2005dq}.
However, it is was only realized very recently that the data can
be understood to depend on only several variables via
the construction of ``scaling relations''~\cite{Adler:2002pu,Back:2004zg}.
These hold despite
the enormous change in the available phase space as the energy
increases, as evidenced by the higher multiplicities and the wider
rapidity distributions.

Calculations performed with ideal hydrodynamics indicate that 
the magnitude of $v_2$ is mainly driven by the eccentricity
of the initial-state matter distribution~\cite{Ollitrault:1992bk}, 
where the eccentricity $\epsilon$ is defined as
\begin{equation}
\epsilon_{std} = \frac{\sigma^2_Y - \sigma^2_X}{\sigma^2_Y + \sigma^2_X} 
\end{equation}
where $\sigma^2_X$ is the variance in the direction along the reaction plane and $\sigma^2_Y$ is the variance perpendicular to it.  

\begin{floatingfigure}{70mm}
\begin{center}
\includegraphics[width=65mm]{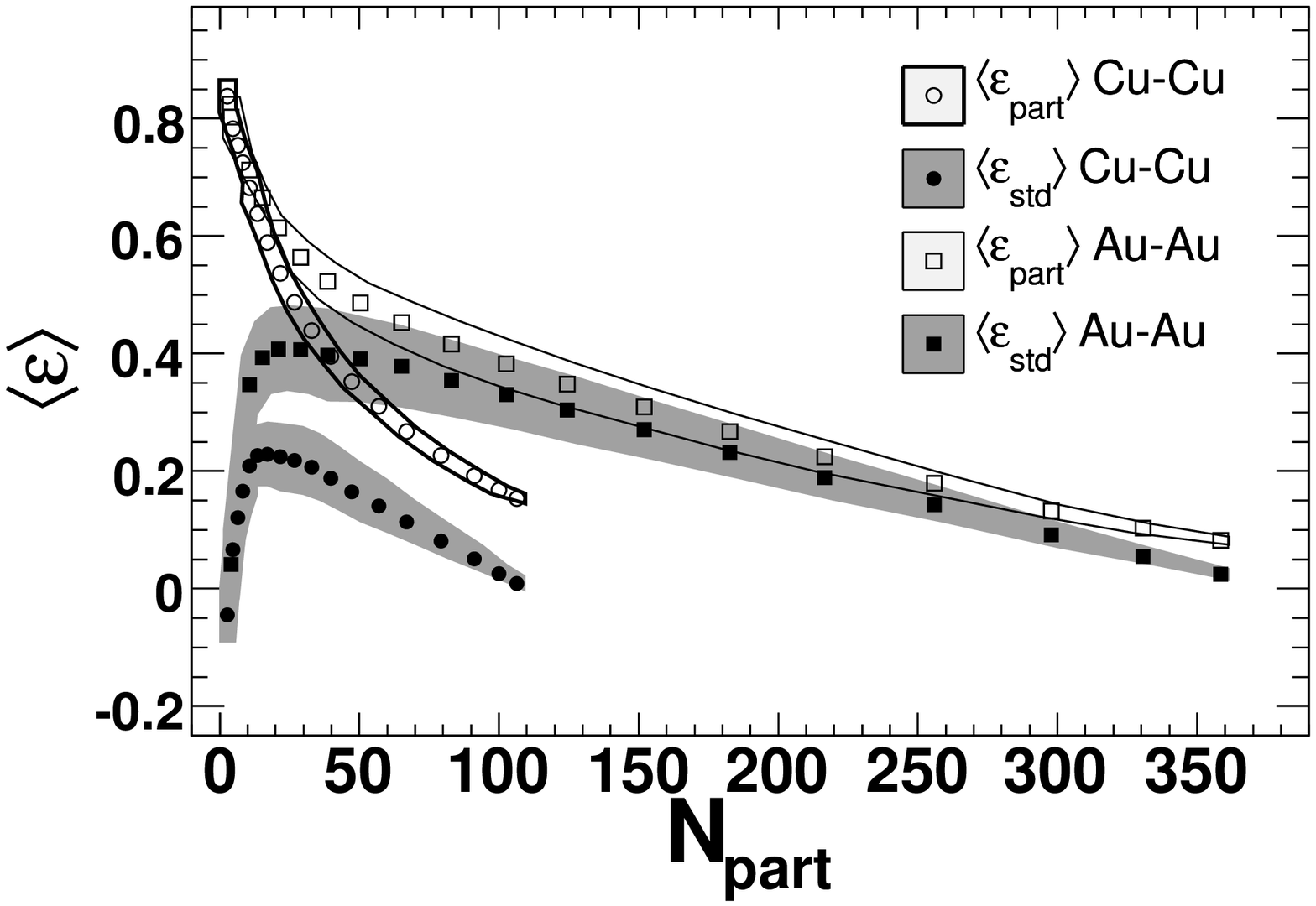}
\includegraphics[width=50mm]{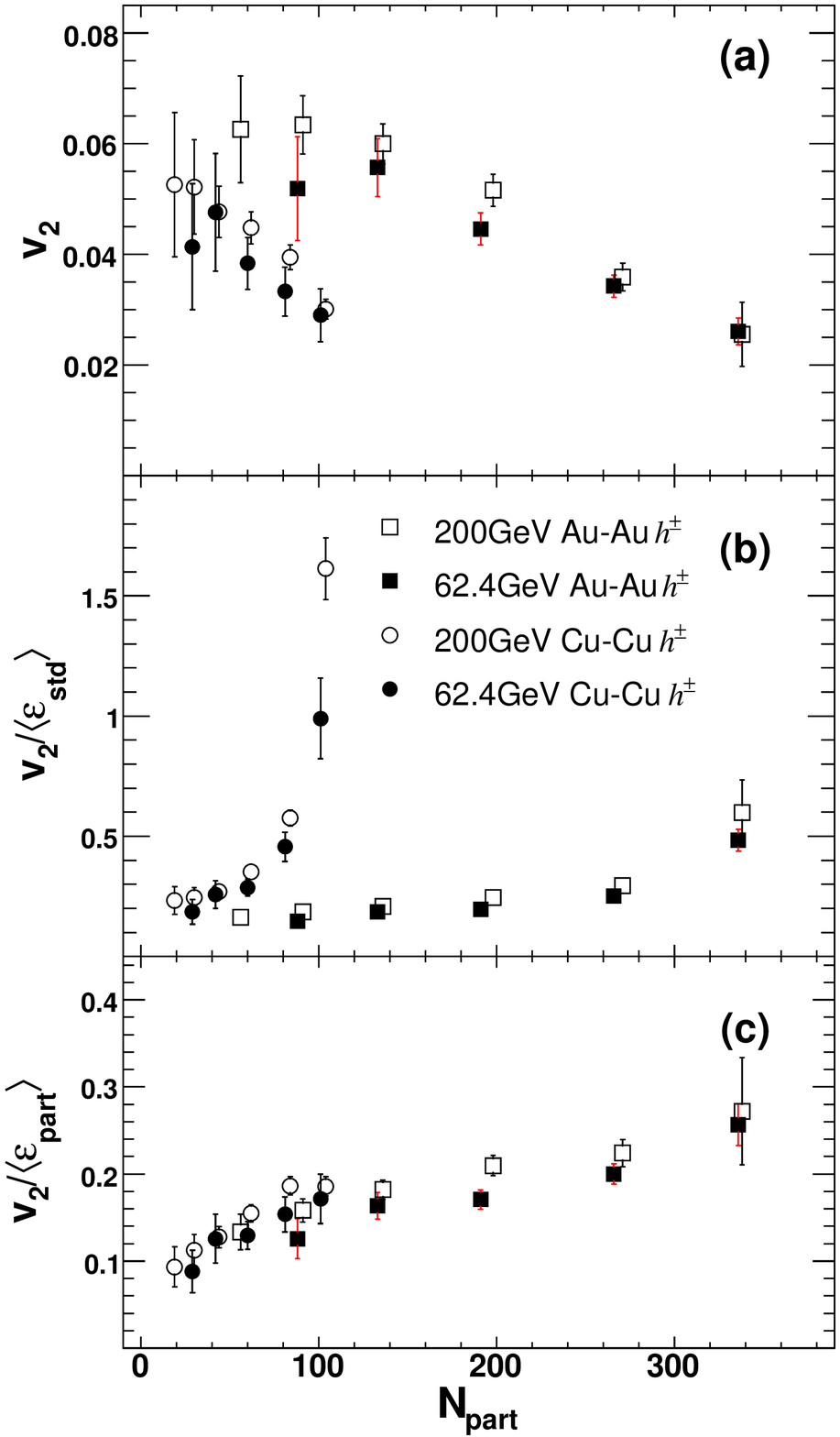}
\caption{
\label{fig:eccpart}
(top) Standard and participant eccentricity for Au+Au and Cu+Cu
collision vs. $\np$.  (bottom) $v_2$, $v_2/\epsilon_{std}$, and
$v_2/\epsilon_{part}$ for Au+Au and Cu+Cu collisions at
200 and 62.4 GeV, from PHOBOS.
}
\end{center}
\end{floatingfigure}

The top panel of Fig.~\ref{fig:eccpart} shows $\epsilon_{std}$ as
a function of $\np$, and illustrates how it goes to 1 for small
$\np$ and large $\np$.
While there is no
principle requiring $v_2$ to be {\it linear} with $\epsilon_{std}$,
it turns out to be in all numerical calculations.
The same hydrodynamic calculations suggest that there is a
maximum value of $v_2/\epsilon$, which has been called the
``hydrodynamic limit'' reflecting that full equilibrium is
reached~\cite{Voloshin:1999gs}.
Again, there is no known deep principle behind
this concept, as it arises out of the interplay of transverse
pressure and longitudinal expansion.  However, all of this
does suggest that $v_2/\epsilon$ is an important diagnostic
variable which removes the ``trivial'' contribution from
the overlap geometry and provides
access to the magnitude of the pressure build-up, and thus
potentially to the equation of state~\cite{Kolb:2003dz}.

Experimental data from several collision energies, and a range
of collision centralities shows that the ``pressure'', as 
diagnosed via $v_2/\epsilon$ is a simple near-linear function
of $dN/dy/\langle S \rangle$, where $S \equiv \pi \sigma_X \sigma_Y$ 
is the transverse area of the initial matter distribution~\cite{Adler:2002pu}.
This
suggests that the pressure is a function of the entropy density
in the transverse plane (if multiplicity is generally thought to 
be linear with the entropy, as will be discussed more later).
Of course, it is observed that the data rises monotonically
with transverse density, or (almost) equivalently with $\np$
as shown for Au+Au in the lower panels of Fig.~\ref{fig:eccpart}.
It appears that the ``hydrodynamic limit'' has been observed yet,
although it has been argued by some authors that
this indicates the {\it approach} to equilibration, rather
than equilibration itself~\cite{Bhalerao:2005mm}.
The current experimental and theoretical situation precludes 
definitive statements either way,
but at the very least it appears to be universal across
changes in both energy and geometry (two ways to the edge of
liquidity, or fluidity).  It also does not seem to be ``broken''
even in very peripheral events or at low energies, where one
might expect the low multiplicities to not favor local
equilibration.

One way to probe the limits of this picture is to thus go
to much smaller systems, Cu+Cu collisions for example, which
reduce the number of participating nucleons by a factor of three.
Data on $v_2$ as a function of $\np$, 
shown in the lower panels of Fig.~\ref{fig:eccpart},
already shows that
these systems appear very different.  The Au+Au data has a
maximum $v_2$ at $\np \sim 100$ and decreases to 1-2\% at
$\np \sim 350$.  Conversely, the Cu+Cu data starts at 
4-5\% and only decreases to 3\% in the most central collisions.
Already this looks very strange, since by construction, 
$\epsilon_{std} \rightarrow 0$ for the most central events.  
How can central Cu+Cu have so much apparent transverse pressure when
there is presumably no geometric way to generate it?  
Things are even more confusing when plotting $v_2/\epsilon_{std}$
as a function of $\np$ for Au+Au and Cu+Cu.  There seems to be
no connection between these two systems.
Theoretical calculations typically assume that the matter
density is a smoothly varying one, described by a Fermi
distribution.%\cite{DeJager:1987qc}.  
However, experiments and many dynamical models
assume that interactions occur by nucleon-nucleon collisions
where individual nucleons are distributed uniformly throughout the
nuclear volume, e.g. as implemented by ``Glauber Monte Carlo''
approaches\cite{Miller:2007ri}.  
The finite number, both in Cu and Au nuclei,
lead to fluctuations in the matter density which are especially
notable in peripheral Au+Au and nearly all Cu+Cu collisions.
If one assumes the matter density is not defined by an
idealized overlap zone of the nuclear densities, but by the
distribution of the participants themselves, 
then the correct scaling should be achieved using
what is called the ``participant eccentricity'' as first
defined by PHOBOS~\cite{Manly:2005zy}
\begin{equation}
\epsilon_{part} = \frac{\sqrt{ (\sigma^2_Y-\sigma^2_X)^2+4(\sigma^2_{XY})^2 }}{\sigma^2_X + \sigma^2_Y}
\end{equation}

\begin{floatingfigure}[l]{60mm}
\begin{center}
\includegraphics[width=55mm]{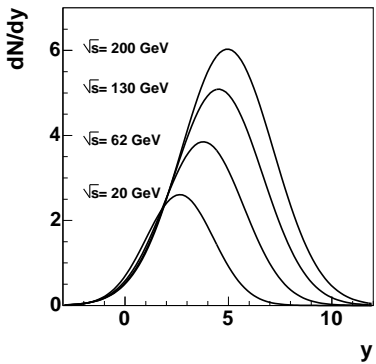}
\caption{
Extended longitudinal scaling in Landau's hydrodynamics,
from Ref.\cite{Steinberg:2004vy}
\label{fig:limfrag_2}
}
\end{center}
\end{floatingfigure}

This measures the shape of the region defined by the principal axes
of the participants, on an event-by-event basis, using a Glauber
Monte Carlo calculation.
This quantity has the advantage over $\epsilon_{std}$ of being
positive definite for all values of impact parameter.  Thus, 
$\epsilon_{part}$ does not necessarily trend to zero for $b \rightarrow 0$,
which is now understood as being due to fluctuations.

Scaling $v_2$ for several collision energies, centralities, and
system sizes by $\epsilon_{part}$ shows a universal scaling
of $v_2/\epsilon_{part}$ with $dN/dy/\langle S \rangle$, or similarly
$\np$ as shown in Fig.~\ref{fig:eccpart}.  
This suggests that the geometrical configuration
of the participants (at least as indicators for the spatial location
of inelastic interactions) is ``frozen in'' immediately, i.e.\ it does
not require substantial time for the deposited energy to locally
equilibrate and react hydrodynamically.  This is of course fully
consistent with the previous estimates of $\tau_0$, but suggests
that there is little room for ``free streaming'' before
thermalization~\cite{Kolb:2003dz}.

\section{Initial Conditions}

Rapid equilibration in strong interactions is a physical scenario with a long
history\cite{Landau:gs,Fermi:1950jd}.  
Landau and Fermi considered this possibility in the
early 1950's when estimating the total multiplicity generated in the
collision of two nuclei, or even nucleons, by
estimating the total entropy assuming all of the incoming
energy is thermalized in a Lorentz-contracted volume.  
In the hydrodynamic paradigm, the strength of the
interactions leaves no natural scales in the problem save two, 
1) the initial longitudinal size and thus the time scale
$\tau_0 \propto \Delta z \propto 1/\gamma \propto 1/\sqrt{s}$,
and the final temperature, or equivalent density, described
above~\cite{Carruthers:dw}.
As described in several recent references (e.g. Ref.~\cite{Steinberg:2004vy}, 
these initial conditions
lead to testable phenomenological consequences.
The multiplicity in the Fermi-Landau approach scales as
$N \propto s^{1/4}$, a power-law behavior that
is compatible with a wide variety of total multiplicity data
from a variety of systems over several orders of magnitude 
in beam energy. It leads to Gaussian rapidity
distributions with variance 
$\sigma^2_y = \ln(\sqrt{s}/2m)$~\cite{Carruthers:dw}.
Combining these two formulae leads to particle densities
that approximately scale when viewed in the rest frame of
one of the projectiles, as a functionn of $\eta - y_{\mathrm{beam}}$,
as shown in Fig.~\ref{fig:limfrag_2}.

\begin{floatingfigure}[r]{50mm}
\begin{center}
\includegraphics[width=45mm]{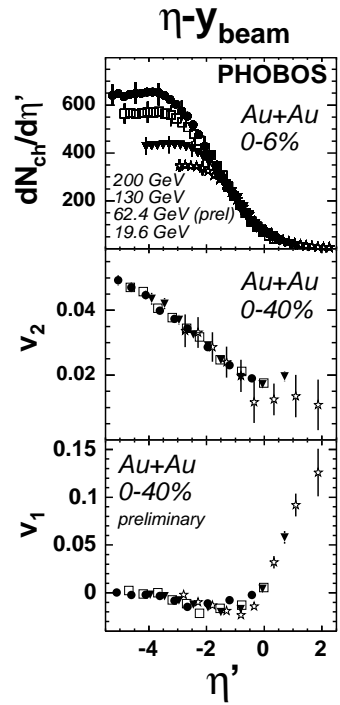}
\caption{
Extended longitudinal scaling seen in PHOBOS data on $dN/d\eta/\np/2$
\label{fig:mult63_lf}
}
\end{center}
\end{floatingfigure}

Landau's initial conditions and dynamical evolution are in some sense 
very different than the one suggested
by Bjorken in the early 1980's, where ``boost-invariance'' was
thought to be the relevant feature of the dynamics of strong
interactions~\cite{Bjorken:1982qr}.  
Boost-invariance is in fact a generic feature of
hydrodynamic solutions, and was noticed by Landau in his hydrodynamic
calculations~\cite{Belenkij:cd}.
In general, the difference between the two is a matter of
evolution time and Landau's initial conditions have a
much different causal structure.  Early thermalization in
the highly compressed initial state leads to strong
longitudinal expansion controlled by $\tau_0$.  Conversely,
boost-invariant initial conditions lead to no longitudinal
pressure gradients, but simply $v=z/t$ (a la like Hubble
expansion).  

RHIC data do not show any clear signs of boost invariance
in the final state.  Intead they have a strong rapidity 
dependence~\cite{Back:2005hs},
also seen by the elliptic flow as a 
function of pseudorapidity~\cite{Back:2004zg}.
Interestingly, they also show an invariance when plotted as
a function of $\eta - y_{beam}$, shown in Fig.~\ref{fig:mult63_lf}
from Ref.~\cite{Roland:2005ei}.
for $dN_{ch}/d\eta$, $v_2$ and $v_1$, circumstantial evidence that
particle production in heavy ion collisions undergoes
early thermalization and thus strong longitudinal expansion.  

It then becomes an interesting question whether all of these
scaling relations indicate a broader validity of the near-perfect
fluid paradigm over most the collision evolution.  The
scaling relations hold as a function of energy,
centrality, and rapidity.  They are thus related to the
question of thermalization time and its effects on longitudinal
dynamics.  

\begin{floatingfigure}[l]{55mm}
\begin{center}
\includegraphics[width=45mm]{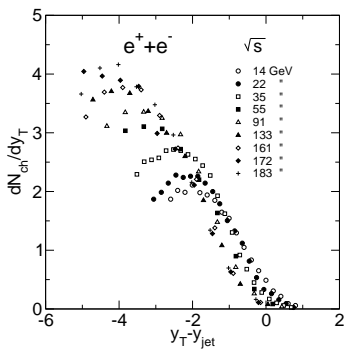}
\caption{Extended longitudinal scaling seen in $\epem$ reactions,
from Ref.~\cite{Back:2004je}.
\label{fig:limfrag_eepp}
}
\end{center}
\end{floatingfigure}

This points directly to an even deeper question, that of whether or not
there is a dynamical length scale in the problem: one that reflects
the microscopic dynamics than simply the boundary conditions.  
If no such scale exists, then there
is no natural way to argue that some systems, e.g. nucleon-nucleon
collisions or $\epem$ annihilation into hadrons, are ``too small'' 
to rapidly thermalize or react hydrodynamically.  Those systems
have been noted to have the same multiplicity 
(i.e.\ entropy)~\cite{Back:2006yw} and
freezeout temperature~\cite{Becattini:1997ii} 
as nominally larger systems, and they
also have been long-known to show extended longitudinal 
scaling~\cite{Back:2004je,Alner:1986xu}.
It is conceivable that all strong interactions could be understood
in part by the near-perfect fluid paradigm.  The criticism that $\epem$
is already understood via perturbative QCD calculations can be
countered by the fact that these calculations seem to show features
that are parametrically similar to Landau's hydrodynamic 
model\cite{Tesima:1989ca,Steinberg:2004wx}.
In the end, all of this will hinge on the viscosity of the produced
matter, something which will be discussed below.

\section{Degrees of Freedom}

And yet, even if the near-perfect fluid paradigm is relevant over
most of the evolution, one major outstanding question remains.
What is the fluid made of?  How did it come into being as a
locally equilibrated state of matter?
\begin{floatingfigure}[r]{80mm}
\begin{center}
\includegraphics[width=60mm]{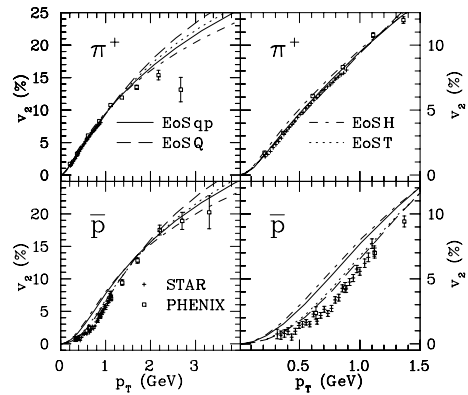}
\caption{
Data on $v_2$ vs. $p_T$ compared to various equations of state, from
Ref.~\cite{Huovinen:2005gy}.
\label{fig:v2piap5}
}
\end{center}
\end{floatingfigure}
Obviously there must be some non-equilibrium dynamics to generate 
the observed entropy.  Nothing in the data 
discussed so far uniquely indentifies which
degrees of freedom are able to achieve this.
The most natural assumption would be that the early stages are
dominated by the dynamics of free quarks and gluons, or at
least the dynamics of quark and gluon fields that are studied
using Lattice QCD.  However, attempts to model the existing
data on $v_2$ vs. $p_T$ for identified particles find that 
a first-order phase transition needs to be put in 
by hand~\cite{Huovinen:2005gy}.
The existing calculations, shown in Fig.~\ref{fig:v2piap5} 
do not provide sufficient ``softening'' of the equation of 
state to model the heavier particles which are most sensitive 
to the speed of sound.

\begin{floatingfigure}{80mm}
\begin{center}
\includegraphics[width=75mm]{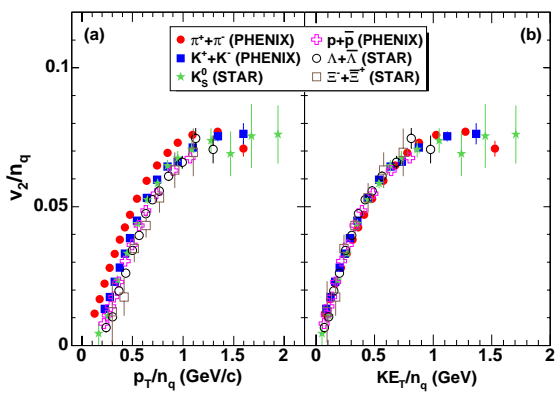}
\caption{
PHENIX data on $v_2$ for different particle species, scaled
by the number of constituent quarks ($n_q$), from Ref.~\cite{Adare:2006ti}.
\label{fig:phenix_cq_Fig3}
}
\end{center}
\end{floatingfigure}

These data have been studied for many different particles species,
which have different mass and quark content.  
Recent PHENIX data~\cite{Adare:2006ti}
shows that all of the available data on $v_2$ vs. $p_T$
lie near one another when plotted as a function of 
$v_2/n_q$ on the Y axis, where $n_q$ is the number of valence
quarks and anti-quarks in the hadron, and $KE_{T}/n_q$ on
the X axis, where $KE_{T} = m_T - m$ for each hadron of mass
$m$.  This suggests a scenario where freezeout occurs by the
recombination of ``constituent quarks'', particles which
have a mass of $\sim m/n_q$ and the right quantum numbers for
each hadron.  However, the same PHENIX paper also suggests
that ``the scaling with valence quark number may indicate
a requirement of a minimum number of objects in a localized
region of space that contain the prerequisite quantum
numbers of the hadron to be formed.  Whether the
scaling further indicates these degrees of freedom are present at the
earliest time is in need of more detailed theoretical
investigation''.  If one considers the overall dynamical
evolution of the system, with many independent
stages in principle~\cite{Steinberg:2004wx}, it is possible that constituent
quark scaling is only probing the very final stages before
formation of the final state hadrons.  Thus, it is premature
to suggest that these RHIC data gives direct information about the
early-time formation of a truly quark and gluon phase.

\begin{floatingfigure}[l]{65mm}
\begin{center}
\includegraphics[width=60mm]{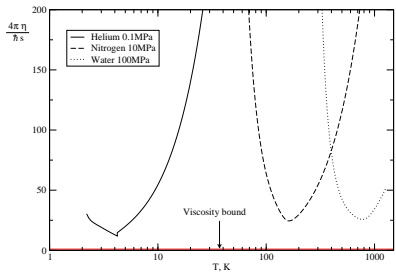}
\caption{
$4\pi\eta/s$ for a variety of systems, with the postulated lower
bound indicated, from Ref.~\cite{Kovtun:2004de}.
\label{fig:graph-He-N-H20}
}
\end{center}
\end{floatingfigure}
Suggesting that quasi-free quarks and gluons are not the active
degrees of freedom at early times does not obviously contradict lattice
data.  In most calculations it is found that
$\epsilon / T^4$, which is proportional to the thermodynamically-active 
number of degrees of
freedom, does not approach the Stefan-Boltzmann limit even
at high temperatures.  In fact it falls short of the limit
by about 20\%.  AdS/CFT-based arguments, which model the
strongly-coupled Yang-Mills plasma as a 10 dimensional black hole, 
predict precisely a 25\% shortfall in the number of degrees
of freedom~\cite{Gubser:1996de}.
Similar arguments have also shown that the shear viscosity
on the gauge theory side is proportional to the entropy 
density, and their ratio has a lower bound~\cite{Kovtun:2004de}, i.e.
%\begin{equation}
$\frac{\eta}{s} \geq \frac{1}{4\pi}$.
%\end{equation}

It is now an active field of investigation to understand the
precise mechanism which implements strong coupling in QCD.
Shuryak, Zahed, Brown and others have proposed the presence
of colored bound states in a QCD plasma~\cite{Shuryak:2004tx, Brown:2004}.  
This leads to larger
cross sections and thus stronger coupling in the system.
And yet, similar lattice calculations, as discussed above, find that
the fluctuations of quark-antiquark fields are not consistent with
the active degrees of freedom being $q\overline{q}$ bound
states~\cite{Majumder:2005ai,Karsch:2005ps}.

Conventional fluids, shown in Fig.~\ref{fig:graph-He-N-H20} 
exceed the postulated viscosity bound by a factor of ten~\cite{Kovtun:2004de}.
The success of the hydrodynamic models shown above suggests
that the viscosity is almost negligible, possibly saturating 
the bound or even being essentially zero.  And yet, the precise value
of $\eta/s$ has not been established using experimental data.
It is clear that saturating the viscosity bound would make an
exciting connection between relativistic heavy ion collisions
and a prediction based on physics nominally far from the
original domain of RHIC physics.

\section{The Future}

\begin{floatingfigure}{70mm}
\begin{center}
\includegraphics[width=65mm]{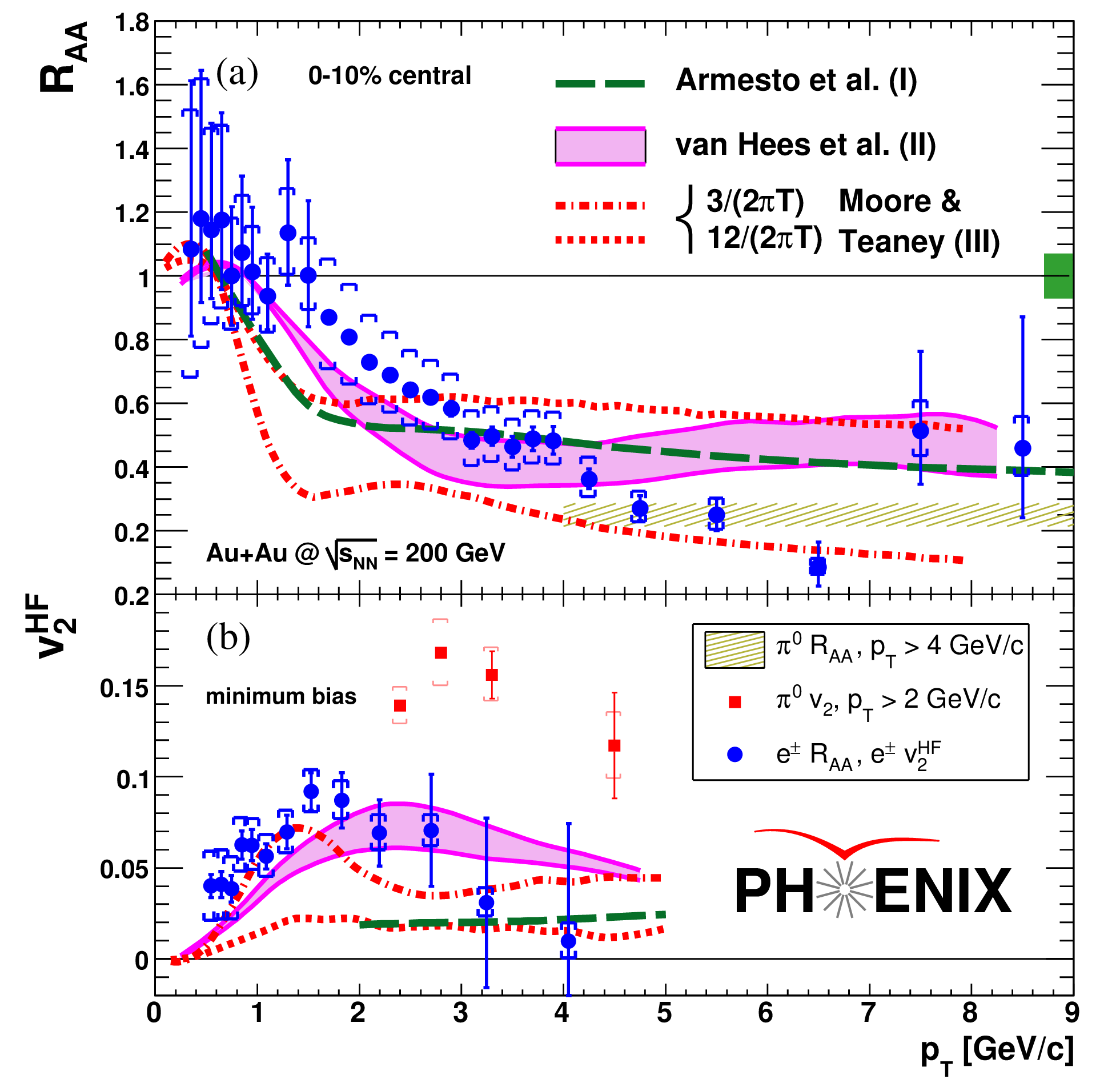}
\caption{
PHENIX data~\cite{Adare:2006nq} on charm suppression and elliptic flow
for Au+Au collisions.
\label{fig:phenix_charm_fig3}
}
\end{center}
\end{floatingfigure}

Probing the transport properties of the system will be the
focus of the next generation of RHIC experiments.  A natural
probe of this is the transport of heavy quarks.
New silicon detectors in PHENIX and STAR are being developed
to measure charmed particles by means of displaced decay
vertices.  By studying the correlation between the flow
and energy loss of charmed mesons, e.g. from recent
PHENIX data~\cite{Adare:2006nq} shown in Fig.~\ref{fig:phenix_charm_fig3}, 
it is possible to
relate their magnitude to a charm diffusion coefficient 
$D$~\cite{Moore:2004tg}.
This in turn allows an estimation of shear viscosity
by the Einstein relation $D \sim 6\eta/sT$.  Already a
first measurement of this quantity has been performed by
PHENIX, which suggests the viscosity seen by charm quarks
falls at most within a factor of 2-3 above the viscosity bound.
The next generation of measurements should substantially improve the
experimental precision.

At the same time, theoretical calculations will also have 
make equivalent strides.  Fully three-dimensional hydrodynamical
calculations will be required, with full control over initial
conditions and freezeout (e.g. Ref.~\cite{Hirano:2004rs}).  
All possible initial conditions,
from Landau's to Bjorken's should be accessible.  This should
be coupled with truly systematic studies in order to assign 
error bars to extracted parameters characterizing 
the initial state, equation of state, and freezeout conditions.
Then we will have a quantitative handle on just how near we are to
the most-perfect fluid.

\ack
The author would like to thank to the GHP organisers for the invitation
to speak in Nashville.
Thanks go
my RHIC colleagues, as always, for useful comments and suggestions
on the talk and proceedings, especially Mark Baker, Wit Busza,
Jamie Nagle,Paul Stankus and Bill Zajc. 
This work was supported in part
by the Office of Nuclear Physics of the U.S. Department of Energy under
contracts: DE-AC02-98CH10886.

\end{document}